\documentclass[notitlepage]{revtex4-1}

\usepackage[utf8]{inputenc}
\usepackage[colorlinks=true,citecolor=blue,linkcolor=blue]{hyperref}
\usepackage[normalem]{ulem}
\usepackage{url}
\usepackage{graphicx,wrapfig,float,slashed,cancel}
\usepackage{amsmath,amssymb,epsfig,graphicx,xcolor,stmaryrd}
\usepackage{bm}
\usepackage{enumitem}

\definecolor{darkblue}{RGB}{1, 90, 173}

\begin{document}


\title{Investigation of full-charm and full-bottom pentaquark states}

\author{K.~Azizi}
\email{ kazem.azizi@ut.ac.ir}
\thanks{Corresponding author}
\affiliation{Department of Physics, University of Tehran, North Karegar Avenue, Tehran
14395-547, Iran}
\affiliation{Department of Physics, Do\v{g}u\c{s} University, Dudullu-\"{U}mraniye, 34775
Istanbul, T\"{u}rkiye}
\affiliation{School of Particles and Accelerators, Institute for Research in Fundamental Sciences (IPM) P.O. Box 19395-5531, Tehran, Iran}
\affiliation{Department of Physics and Technical Sciences, Western Caspian University, Baku AZ 1001, Azerbaijan}
\author{Y.~Sarac}
\email{yasemin.sarac@atilim.edu.tr}
\affiliation{Electrical and Electronics Engineering Department,
Atilim University, 06836 Ankara, T\"{u}rkiye}
\author{H.~Sundu}
\email{ hayriyesundu.pamuk@medeniyet.edu.tr}
\affiliation{Department of Physics Engineering, Istanbul Medeniyet University, 34700 Istanbul, T\"{u}rkiye}

\date{\today}

\preprint{}

\begin{abstract}

The continuous advancement of experimental techniques and investigations has led to observations of various exotic states in particle physics. Each addition to this family of states not only raises expectations for future discoveries but also focuses attention on such  potential new states. Building upon this motivation and inspired by recent observations of various traditional and  exotic particles containing an increased number of heavy quarks, our study provides a spectroscopic search for potential pentaquark states with spin-parity $\frac{3}{2}^-$ and composed entirely of charm or bottom quarks. We predict the masses for full-charm and full-bottom pentaquark states as $m = 7628 \pm 112$~MeV and $m = 21982 \pm 144$~MeV, respectively.  We also compute the  current couplings of these states to vacuum, which are main inputs in investigations of their various possible decays. 
   
\end{abstract}


\maketitle

\renewcommand{\thefootnote}{\#\arabic{footnote}}
\setcounter{footnote}{0}
\section{\label{sec:level1}Introduction}\label{intro} 

Since the proposal of the quark model, hadrons with non-conventional structures, which do not fit the conventional baryons composed of three quarks (antiquarks) and mesons composed of a quark and an antiquark, have been subjects of interests. The theory of strong interaction does not rule out the existence of such states, and this has attracted  interest in these states. They were investigated extensively in both theory and experiments. Finally, the first evidence came out with the observation of $X(3872)$ state in 2003~\cite{Choi:2003ue}. And following this observation, many other such exotic state candidates were observed~\cite{Zyla:2020zbs,Aaij:2015tga,Aaij:2016ymb,Aaij:2019vzc,LHCb:2021auc,LHCb:2021vvq,LHCb:2021chn,LHCb:2020jpq,LHCb:2022ogu} and listed in Particle Data Group (PDG)~\cite{Workman:2022ynf}. These observations were also followed by many theoretical investigations trying to explain their internal structures, which still have ambiguity and need to be clearly identified with more scrutiny. It is evident that we will come across with other such possible exotic states in the future. This expectation necessitates their examinations in detail via different approaches to provide an understanding of their substructure and properties and provide feedback for future  investigations. Besides, these states help to deepen our understanding of the dynamics of the strong interaction. 

The pentaquark states are among these non-conventional states with their first observation reported in 2015 by the LHCb collaboration~\cite{Aaij:2015tga}. The investigation of the $\Lambda_b^0 \rightarrow J/\psi p K^-$ process resulted in two pentaquark states in the $J/\psi p$ invariant mass spectrum with the following resonance parameters~\cite{Aaij:2015tga}: $m_{P_c(4380)^+}=4380 \pm8 \pm 29~\mathrm{MeV}$, $\Gamma_{P_c(4380)^+}=205 \pm 18 \pm 86~\mathrm{MeV}$ and $m_{P_c(4450)^+}=4449.8 \pm 1.7 \pm 2.5~\mathrm{MeV}$, $\Gamma_{P_c(4450)^+}= 39 \pm 5 \pm 19~\mathrm{MeV}$. This observation was supported by a full amplitude analysis for $\Lambda_b^0 \rightarrow J/\psi p \pi^-$ decays~\cite{Aaij:2016ymb} in 2016. In 2019, using updated data, a new pentaquark state, $P_c(4312)^+$, was reported with $m_{P_c(4312)^+}=4311.9 \pm 0.7^{ +6.8}_{-0.6}~\mathrm{MeV}$ and $\Gamma_{P_c(4312)^+}=9.8 \pm 2.7 ^{ +3.7}_{-4.5}~\mathrm{MeV}$ by the LHCb collaboration and analyses revealed two narrow overlapping peaks for the previously observed peak of the $P_c(4450)^-$ state with masses and widths: $m_{P_c(4440)^+}=4440.3 \pm 1.3 ^{+ 4.1}_{-4.7}~\mathrm{MeV}$, $\Gamma_{P_c(4440)^+}= 20.6 \pm 4.9^{+8.7}_{-10.1}~\mathrm{MeV}$ and $m_{P_c(4457)^+}=4457.3 \pm 0.6 ^{+ 4.1}_{-1.7}~\mathrm{MeV}$, $\Gamma_{P_c(4457)^+}= 6.4 \pm 2.0^{+5.7}_{-1.9}~\mathrm{MeV}$~\cite{Aaij:2019vzc}. In the recent investigations, new states with the strange quark were also added to this family. The LHCb collaboration reported the $P_{cs}(4459)^0$ state through the investigation of $J\psi \Lambda$ invariant mass distribution in $\Xi_b^-\rightarrow J/\psi K^-\Lambda$ decays~\cite{LHCb:2020jpq}. The mass and width for the $P_{cs}(4459)^0$ were given as $m=4458.8 \pm 2.9^{+4.7}_{-1.1}~\mathrm{MeV}$, and $\Gamma = 17.3 \pm 6.5^{+8.0}_{-5.7}~\mathrm{MeV}$~\cite{LHCb:2020jpq}. $P_{cs}(4338)$ state was reported with the mass $4338.2 \pm 0.7\pm 0.4$~MeV and the width $ 7.0\pm 1.2 \pm 1.3$~MeV from the amplitude analyses of $B^{-}\rightarrow J/\psi \Lambda \bar{p}$~\cite{LHCb:2022ogu}.

Following the observations of the above pentaquark states, theoretical researches chasing the purpose of identifying their various properties have focused on these states. The sub-structures and quantum numbers of these observed pentaquark states still have uncertainty and this makes them attractive theoretically. Their investigations provide information not only about their nature and substructure but also for future experiments searching for new such states. Their scrutiny also supports and contributes to improving our understanding of the nonperturbative domain of quantum chromodynamics (QCD) with their distinct quark substructures compared to conventional hadrons with three quarks/antiquarks or a quark and an antiquark. With all these issues, these states were investigated thoroughly with the application of different approaches assigning them different substructures such as diquark-diquark-antiquark~\cite{Lebed:2015tna,Li:2015gta,Maiani:2015vwa,Anisovich:2015cia,Wang:2015ava,Wang:2015epa,Wang:2015ixb,Ghosh:2015ksa,Wang:2015wsa,Zhang:2017mmw,Wang:2019got,Wang:2020rdh,Ali:2020vee,Wang:2016dzu,Wang:2020eep} and diquark-triquark~\cite{Wang:2016dzu,Zhu:2015bba} models, and meson baryon molecular states ~\cite{Chen:2015loa,Chen:2015moa,He:2015cea,Meissner:2015mza,Roca:2015dva,Azizi:2016dhy,Azizi:2018bdv,Azizi:2020ogm,Chen:2020opr,Li:2024wxr,Li:2024jlq,Wang:2023eng,Chen:2020uif,Peng:2020hql,Chen:2020kco,Wang:2019hyc,Wang:2021itn,Wang:2018waa}. Topological soliton model~\cite{Scoccola:2015nia} and a variant of the D4-D8 model~\cite{Liu:2017xzo} were applied to search their properties, and in Refs~\cite{Guo1,Guo2,Mikhasenko:2015vca,Liu1,Bayar:2016ftu} their being kinematical effects were taken into account. With the possibility for observation of new pentaquark states with quark content different from the observed ones, various candidate pentaquark states were also studied in the literature~\cite{Liu:2020cmw,Chen:2015sxa,Feijoo:2015kts,Lu:2016roh,Irie:2017qai,Chen:2016ryt,Zhang:2020cdi,Paryev:2020jkp,Gutsche:2019mkg,Azizi:2017bgs,Cao:2019gqo,Azizi:2018dva,Zhang:2020vpz,Wang:2020bjt,Xie:2020ckr}.

These observations and the advances in experimental facilities and techniques indicate the possibility of observing more exotic states in the future. With this expectation, these states with different possible structures and quark contents have been studied using various models~\cite{Liu:2020cmw,Chen:2015sxa,Feijoo:2015kts,Lu:2016roh,Irie:2017qai,Chen:2016ryt,Zhang:2020cdi,Paryev:2020jkp,Gutsche:2019mkg,Azizi:2017bgs,Cao:2019gqo,Azizi:2018dva,Zhang:2020vpz,Wang:2020bjt,Xie:2020ckr,Duan:2024uuf,Kong:2024scz,Sharma:2024wpc,Zhang:2023teh,Yan:2023iie,Liu:2023clr,Wang:2023mdj,Wang:2023ael,Yang:2023dzb,Liu:2023oyc,Paryev:2023uhl,Lin:2023iww,Xin:2023gkf,Chen:2023qlx,Zhu:2023hyh,Yan:2023kqf}. The observed traditional baryons with double valence heavy quarks or exotic states with full heavy quarks in the recent years~\cite{LHCb:2017iph,LHCb:2018pcs,LHCb:2020bwg,Bouhova-Thacker:2022vnt,Zhang:2022toq,CMS:2023owd,ATLAS:2023bft} enhance the expectation for future observation of such pentaquark states with two, three, or more heavy valence quarks in their quark substructures. This expectation motivates us to study such types of systems, and to this end, in this work, our focus is to study the pentaquark stets with full charm or full bottom quark contents.  With the same motivation, the pentaquark states with full charm or bottom compositions were also studied in Refs.~\cite{Zhang:2020vpz,Wang:2021xao,Yan:2021glh,An:2020jix,An:2022fvs,Zhang:2023hmg,Yang:2022bfu,Liang:2024met}. However, to enhance our knowledge, better understand these states, and give support for possible future experiments, we need more analysis to explain their nature and possible structures. Furthermore, such analysis helps improve our understanding of nonperturbative regime of the quantum chromodynamics and refines our knowledge about the dynamics of heavy quarks. To investigate these states, we apply a method that has revealed its success so far with its predictions consistent  with the experimental observations, namely the QCD sum rule method~\cite{Shifman:1978bx,Shifman:1978by,Ioffe81}. Using this method, we aim to calculate the masses and current coupling constants for considered states with spin-parity $\frac{3}{2}^-$ with a proper choice of interpolating current defining these states and composed of heavy quark fields. The information obtained about such states, which have the potential to be observed and investigated in future experiments, may provide support to conduct  these experiments and help to explore and understand their properties.  

The outline of the paper is as follows: In Section~\ref{II}, the QCD sum rules giving the masses and current coupling constants of the considered states were obtained. Section~\ref{III} gives the numerical analyses of the QCD sum rule results attained. The last section is devoted to the summary and conclusion.

\section{The QCD sum rules for the masses of full-charm and full-bottom pentaquarks}\label{II}

This section provides the QCD sum rule calculations that give the masses and corresponding current coupling constants of the considered full-charm and full-bottom pentaquarks. The mass calculations with the QCD sum rule method start with a two-point correlation function given as 
\begin{equation}
\Pi_{\mu\nu} (p)=i\int d^{4}xe^{ip\cdot
x}\langle 0|\mathcal{T} \{\eta_{\mu}^{P_{(5Q)}}(x)\bar{\eta}_{\nu}^{P_{(5Q)}}(0)\}|0\rangle,
\label{eq:CorrF1PQ}
\end{equation}
where $\mathcal{T}$ stands for time ordering operator and $\eta_{\mu}^{P_{(5Q)}}$ represents the full-heavy pentaquark currents either composed of  completely charm quarks or completely bottom ones with the following form:
\begin{eqnarray}
\eta^{P_{(5Q)}}_{\mu}&=&[\epsilon^{ijk} Q^{T}_{i}C\gamma_{\mu} Q_{j} Q_{k}][\bar{Q}_{l} i\gamma_5  Q_{l}].\label{Current}
\end{eqnarray}
The $Q$ in Eq.~(\ref{Current}) is either charm quark or bottom quark field, $T$ represents transpose,  $i,~j,~k,~l$ are the color indices, and $C$ is the charge conjugation operator. We use this  interpolating current to investigate  both the fully-charmed pentaquark state using all its heavy quark fields as $c$ quark fields and the fully-bottom pentaquark state using them as all $b$ quark fields.

In the calculation of Eq.~(\ref{eq:CorrF1PQ}), two paths are followed; in the first path, it is calculated in terms of hadronic degrees of freedom. Since the result of the calculation contains the mass of the hadron and the current coupling constant, this path is called the hadronic side. On the other hand, the same equation is also computed in terms of QCD degrees of freedom, that is, in terms of the masses of the included quarks, quark-gluon condensates, and QCD coupling constant. Therefore, this path is called the QCD side of the calculation. After following these two paths to study the same correlation function, the obtained results are matched considering the coefficients of the same Lorentz structure on each part using a dispersion relation to obtain the QCD sum rules of the required quantities, masses, and current coupling constants in this work. The obtained results also take contributions from higher states and continuum, which need to be suppressed to obtain the required physical parameters. Borel transformations and continuum subtractions are applied to both sides to provide this. 

On the hadronic side, the interpolating currents are treated as annihilation or creation operators of the considered hadrons. A complete set of hadronic states carrying the same quantum numbers as the interpolating current is inserted inside the correlation function between the interpolating currents, and the integral over four-$x$ is performed. With these operations, the result becomes
\begin{eqnarray}
\Pi^{\mathrm{Had}}_{\mu\nu}(p)= \frac{\langle 0|J_{\mu}|P_{(5Q)}(p,s)\rangle \langle P_{(5Q)}(p,s)|\bar{J}_{\nu}|0\rangle}{m_{P_{(5Q)}}^2-p^2}+\cdots.
\label{eq:hadronicside1}
\end{eqnarray}
In Eq.~(\ref{eq:hadronicside1}) the $\cdots$ is used for the contribution of higher states and continuum, $|P_{(5Q)}(p,s)\rangle$ is one particle satate with momentum $p$ and spin $s$, and the matrix elements are defined in terms of the current coupling constants, $\lambda_{P_{(5Q)}}$, and Rarita-Schwinger spinor, $u_{\mu}(p,s)$, as 
\begin{eqnarray}
\langle 0|J_{\mu}|P_{(5Q)}(p,s)\rangle &=& \lambda_{P_{(5Q)}} u_{\mu}(p,s).
\label{eq:matrixelement}
\end{eqnarray}
Though the states analyzed here are spin-$\frac{3}{2}$ states, the current used in the calculation also couples to the spin-$\frac{1}{2}$ state. Therefore, to isolate the contribution to the analyses from the spin-$\frac{3}{2}$ state and remove that from spin-$\frac{1}{2}$ state, we need to consider the matrix element entering the calculation relevant for the spin-$\frac{1}{2}$ states as
\begin{eqnarray}
\langle 0|J_{\mu}|\frac{1}{2}(p,s)\rangle =A_{\frac{1}{2}^+}(\gamma_{\mu}+\frac{4p_{\mu}}{m_{\frac{1}{2}^+}})\gamma_5 u(q,s),
\end{eqnarray}
which indicates that the results getting contributions from these states are proportional to $\gamma_{\mu}$ or $p_{\mu}$. According to these, if we choose a proper Lorentz structure that does not contain $\gamma_{\mu}$ or $p_{\mu}$, we separate the part of the results giving contribution only to spin-$\frac{3}{2}$ state. To this end, we take into account the Lorentz structure $g_{\mu\nu}$ in our computation. After using the matrix element given in Eq.~(\ref{eq:matrixelement}) inside the Eq.~(\ref{eq:hadronicside1}), we use the following summation over spin:
\begin{eqnarray}\label{Rarita}
\sum_s  u_{\mu} (p,s)  \bar{u}_{\nu} (p,s) &= &-(\!\not\!{p} + m)\Big[g_{\mu\nu} -\frac{1}{3} \gamma_{\mu} \gamma_{\nu} - \frac{2p_{\mu}p_{\nu}}{3m^{2}} +\frac{p_{\mu}\gamma_{\nu}-p_{\nu}\gamma_{\mu}}{3m} \Big],
\end{eqnarray}
and obtain the hadronic side as
\begin{eqnarray}\label{PhyssSide}
\Pi_{\mu\nu}^{\mathrm{Had}}(p)&=&\frac{\lambda_{P_{(5Q)}}{}^{2}}{p^{2}-m_{P_{(5Q)}}^{2}}(\!\not\!{p} + m_{P_{(5Q)}})\Big[g_{\mu\nu} -\frac{1}{3} \gamma_{\mu} \gamma_{\nu} - \frac{2p_{\mu}p_{\nu}}{3m_{P_{(5Q)}}^{2}} +\frac{p_{\mu}\gamma_{\nu}-p_{\nu}\gamma_{\mu}}{3m_{P_{(5Q)}}} \Big]+\cdots.
\end{eqnarray}
As stated, to avoid the pollution of the spin-$\frac{1}{2}$ state, we work with the coefficient of the structure $g_{\mu\nu}$, which leads us to 
\begin{eqnarray}\label{PhyssSide}
\Pi_{\mu\nu}^{\mathrm{Had}}(p)&=&\frac{\lambda_{P_{(5Q)}}^{2}}{p^{2}-m_{P_{(5Q)}}^{2}}m_{P_{(5Q)}}g_{\mu\nu} +\cdots,
\end{eqnarray}
where the contributions of other structures are represented by $\cdots$. After the Borel transformation with respect to $-p^2$ this result turns into
\begin{eqnarray}
\Pi_{\mu\nu}^{\mathrm{Had}}(p)&=&\lambda_{P_{(5Q)}}^{2} m_{P_{(5Q)}}e^{-\frac{m_{P_{(5Q)}}^{2}}{M^2}}g_{\mu\nu} +\cdots,
\label{PhyssSideF}
\end{eqnarray}
where $M^2$ is the Borel parameter arising from the Borel transformation.

After getting the hadronic side, we turn our attention to the QCD side of the calculation and use the same correlation function, Eq.~(\ref{eq:CorrF1PQ}), with the explicit form of the interpolating current given in terms of the quark fields. Using Wick's theorem, we obtain the possible contractions between the quark fields and attain the result in terms of the heavy quark propagators as:
\begin{eqnarray}
\Pi_{\mu\nu}^{\mathrm{QCD}}(p)&=&i\int d^4x e^{ip\cdot x}2\epsilon_{abc}\epsilon_{a'b'c'}\Big\{\mathrm{Tr}[\gamma_{\nu}\tilde{S}_Q^{ba'}(x) \gamma_{\mu}S_Q^{ab'}(x)]\mathrm{Tr}[i \gamma_{5} S_Q^{ee'}(x) i \gamma_{5}S_Q^{e'e}(-x)] S_Q^{cc'}(x)\nonumber\\
&-&\mathrm{Tr}[\gamma_{\nu}\tilde{ S}_Q^{ba'}(x) \gamma_{\mu}S_Q^{ab'}(x)] S_Q^{ce'}(x) i \gamma_{5} S_Q^{e'e}(-x)  i \gamma_{5}S_Q^{ec'}(x)\nonumber\\
&-&2\mathrm{Tr}[i\gamma_{5}S_Q^{ee'}(x) i \gamma_{5}S_Q^{e'e}(-x)] S_Q^{ca'}(x)  \gamma_{\nu}  \tilde{S}_Q^{ab'}(x)   \gamma_{\mu}S_Q^{bc'}(x)\nonumber\\
&+&2S_Q^{ce'}(x)i \gamma_{5} S_Q^{e'e}(-x) i \gamma_{5} S_Q^{ea'}(x)  \gamma_{\nu}  \tilde{S}_Q^{ab'}(x)   \gamma_{\mu}S_Q^{bc'}(x)\nonumber\\
&+&2S_Q^{ca'}(x) \gamma_{\nu} \tilde{S}_Q^{ab}(x)  \gamma_{\mu} S_Q^{be'}(x) i \gamma_{5} S_Q^{e'e}(-x)   i \gamma_{5}S_Q^{ec'}(x)\nonumber\\
&-&2 \mathrm{Tr}[S_Q^{ea'}(x) \gamma_{\nu}\tilde{S}_Q^{ab'}(x) \gamma_{\mu}S_Q^{be'}(x) i\gamma_{5} S_Q^{e'e}(-x) i \gamma_{5}] S_Q^{cc'}(x) \nonumber\\
&+&2S_Q^{cb'}(x) \gamma_{\nu} \tilde{S}_Q^{ea'}(x) i \gamma_{5} \tilde{S}_Q^{e'e}(-x) i \gamma_{5} \tilde{S}_Q^{be'}(x)   \gamma_{\mu}S_Q^{ac'}(x)\Big\},
\label{eq:QCDSide}
\end{eqnarray}
where $\tilde{S}_Q^{ab}(x)=C S_Q^{abT}(x)C$ and  $S_Q(x)$ is the heavy quark propagator given as 
\begin{eqnarray}
S_{Q,{ab}}(x)&=&\frac{i}{(2\pi)^4}\int d^4k e^{-ik \cdot x} \left\{
\frac{\delta_{ab}}{\!\not\!{k}-m_Q}
-\frac{g_sG^{\alpha\beta}_{ab}}{4}\frac{\sigma_{\alpha\beta}(\!\not\!{k}+m_Q)+
(\!\not\!{k}+m_Q)\sigma_{\alpha\beta}}{(k^2-m_Q^2)^2}\right.\nonumber\\
&&\left.+\frac{\pi^2}{3} \langle \frac{\alpha_sGG}{\pi}\rangle
\delta_{ij}m_Q \frac{k^2+m_Q\!\not\!{k}}{(k^2-m_Q^2)^4}
+\cdots\right\}.
\label{eq:Qpropagator}
\end{eqnarray}
Using this propagator in Eq.~(\ref{eq:QCDSide}), and taking the integral over $x$ and applying the Borel transformation, we obtain the final results, which are long and not presented here, given in the following form:
\begin{eqnarray}
\mathcal{B}\Pi^{\mathrm{QCD}}(s_0,M^2)=\int_{25m_Q^2}^{s_0} ds e^{-\frac{s}{M^2}}\rho(s),
\label{Eq:Cor:QCD}
\end{eqnarray}
with $\rho(s)$ being the result corresponding to the coefficient of the Lorentz structure $g_{\mu\nu}$ obtained in the QCD side of the calculation.

The match of the results of the coefficients of the Lorentz structure $g_{\mu\nu}$ attained in hadronic and QCD sides gives the QCD sum rule for the masses and current coupling constants as
\begin{eqnarray}
\lambda_{P_{(5Q)}}^{2} m_{P_{(5Q)}}e^{-\frac{m_{P_{(5Q)}}^{2}}{M^2}}=\mathcal{B}\Pi^{\mathrm{QCD}}(s_0,M^2),
\label{QCDsumrule}
\end{eqnarray}
from which we obtain the mass by taking the derivative of the Eq.~(\ref{QCDsumrule}) with respect to $-\frac{1}{M^2}$ and dividing this new result by Eq.~(\ref{QCDsumrule}) itself. This leads us to the mass as
\begin{eqnarray}
m_{P_{(5Q)}}^2=\frac{\frac{d}{d(-\frac{1}{M^2})}\mathcal{B} \Pi^{\mathrm{QCD}}(s_0,M^2)}{\mathcal{B}\Pi^{\mathrm{QCD}}(s_0,M^2)}, 
\end{eqnarray}  
and current coupling constant is obtained from Eq.~(\ref{QCDsumrule})using the mass result as
\begin{eqnarray}
\lambda_{P_{(5Q)}}^2=\frac{e^{\frac{m_{P_{(5Q)}}^2}{M^2}}}{m_{P_{(5Q)}}}\mathcal{B}\Pi^{\mathrm{QCD}}(s_0,M^2).
\end{eqnarray}
The next stage is to analyze these results with the proper input parameters and fix the corresponding numerical values for these quantities, as explained in the next section.

\section{Numerical Analyses}\label{III}

\begin{figure}[h!]
\begin{center}
\includegraphics[totalheight=5cm,width=7cm]{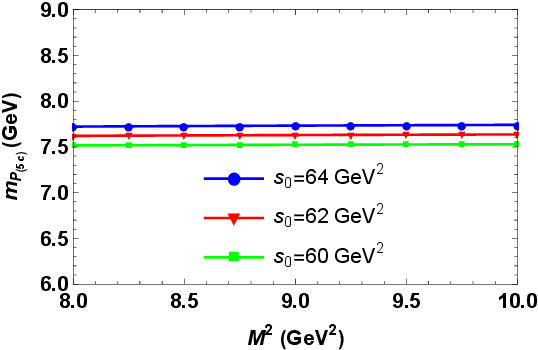}
\includegraphics[totalheight=5cm,width=7cm]{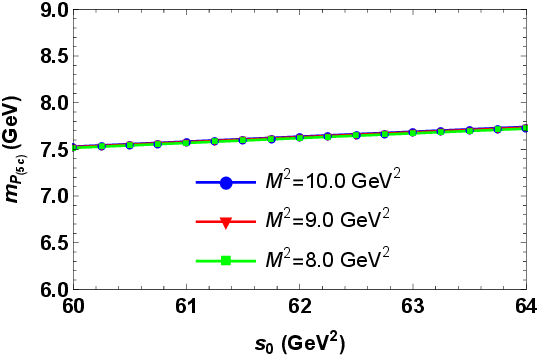}
\end{center}
\caption{\textbf{Left:} The variation of the mass of the $P_{(5c)}$ state as a function of the Borel parameter $M^2$ at various values of the threshold parameter $s_0$. \textbf{Right:} The variation of the mass of the $P_{(5c)}$ state as a function of the threshold parameter $s_0$ at various values of the Borel parameter $M^2$.}
\label{gr:1}
\end{figure}

\begin{figure}[h!]
\begin{center}
\includegraphics[totalheight=5cm,width=7cm]{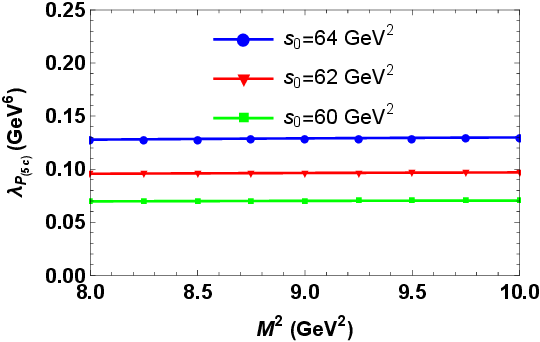}
\includegraphics[totalheight=5cm,width=7cm]{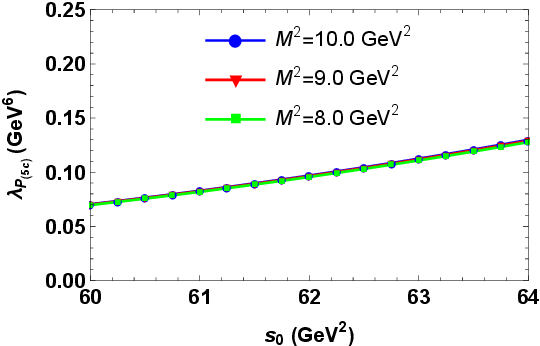}
\end{center}
\caption{\textbf{Left:} The variation of the current coupling constant $\lambda_{P_{(5c)}}$ of the $P_{(5c)}$ state as a function of the Borel parameter $M^2$ at various values of the threshold parameter $s_0$. \textbf{Right:} The variation of the current coupling constant $\lambda_{P_{(5c)}}$ of the $P_{(5c)}$ state as a function of the threshold parameter $s_0$ at various values of the Borel parameter $M^2$.}
\label{gr:2}
\end{figure}

\begin{figure}[h!]
\begin{center}
\includegraphics[totalheight=5cm,width=7cm]{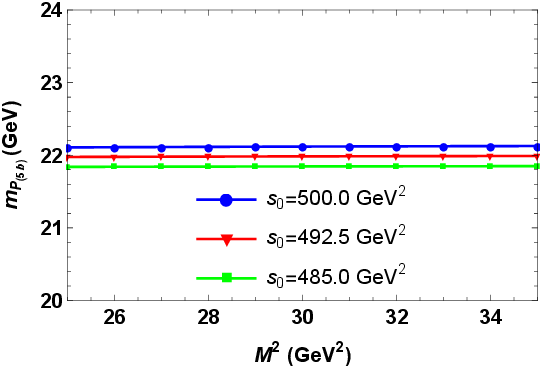}
\includegraphics[totalheight=5cm,width=7cm]{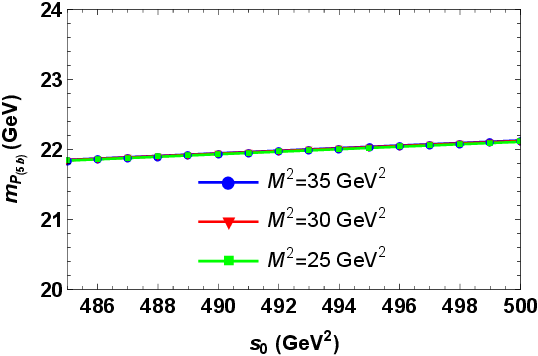}
\end{center}
\caption{\textbf{Left:} The variation of the mass of the $P_{(5b)}$ state as a function of the Borel parameter $M^2$ at various values of the threshold parameter $s_0$. \textbf{Right:} The variation of the mass of the $P_{(5b)}$ state as a function of the threshold parameter $s_0$ at various values of the Borel parameter $M^2$.}
\label{gr:3}
\end{figure}

\begin{figure}[h!]
\begin{center}
\includegraphics[totalheight=5cm,width=7cm]{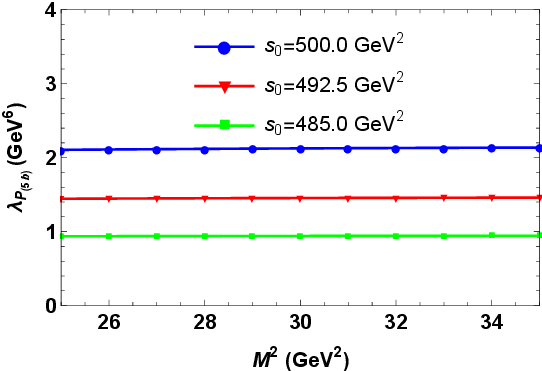}
\includegraphics[totalheight=5cm,width=7cm]{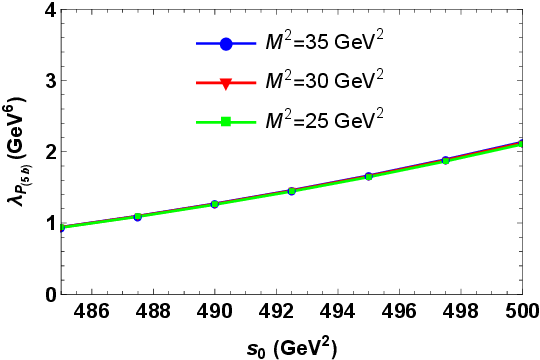}
\end{center}
\caption{\textbf{Left:} The variation of the current coupling constant $\lambda_{P_{(5b)}}$ of the $P_{(5b)}$ state as a function of the Borel parameter $M^2$ at various values of the threshold parameter $s_0$. \textbf{Right:} The variation of the current coupling constant $\lambda_{P_{(5b)}}$ of the $P_{(5b)}$ state as a function of the threshold parameter $s_0$ at various values of the Borel parameter $M^2$.}
\label{gr:4}
\end{figure}

To attain the numerical results from the QCD sum rules obtained for masses and current coupling constants, some input parameters are required. These input parameters are provided in Table~\ref{tab:Inputs}.  
\begin{table}[h!]
\begin{tabular}{|c|c|}
\hline\hline
Parameters & Values \\ \hline\hline
$m_{c}$                                    & $1.27\pm 0.02~\mathrm{GeV}$ \cite{Workman:2022ynf}\\
$m_{b}$                                     & $4.18^{+0.03}_{-0.02}~\mathrm{GeV}$ \cite{Workman:2022ynf}\\
$\langle \frac{\alpha_s}{\pi} G^2 \rangle $ & $(0.012\pm0.004)$ $~\mathrm{GeV}^4 $\cite{Belyaev:1982cd}\\
\hline\hline
\end{tabular}%
\caption{Necessary input parameters for the numerical analyses.}
\label{tab:Inputs}
\end{table} 
Besides, we need two auxiliary parameters in each case. These auxiliary parameters are Borel masses $M^2$ and threshold parameters $s_0$ that enter the calculations by the applications of Borel transformation and continuum subtraction, respectively. In the analyses, the working windows for these parameters are fixed following the standard criteria of the QCD sum rule application. To determine the upper limits of the Borel parameters, dominance of the pole over the continuum is demanded, and for the lower limits, the convergence of the OPE calculation is taken into account. Regarding these, the proper intervals for the Borel parameters are fixed as follows:
\begin{eqnarray}
8.0~\mbox{GeV}^2\leq M^2\leq 10.0~\mbox{GeV}^2,
\end{eqnarray}
for full-charm $P_{(5c)}$ pentaquark state and
\begin{eqnarray}
25.0~\mbox{GeV}^2\leq M^2\leq 35.0~\mbox{GeV}^2,
\end{eqnarray}
for the full-bottom $P_{(5b)}$ pentaquark state. Regarding the intervals of the threshold parameters, these parameters are not completely arbitrary and have connections to the energies of the possible excited states. For this reason, their values are fixed by conjecturing the probable energies that may correspond to the first excited states of the considered pentaquarks. Therefore, the following intervals are applied in the analyses:
\begin{eqnarray}
&60.0~\mbox{GeV}^2 \leq s_0 \leq 64.0~\mbox{GeV}^2 &
\end{eqnarray} 
for $P_{(5c)}$ state and 
\begin{eqnarray}
&485.0~\mbox{GeV}^2 \leq s_0 \leq 500.0~\mbox{GeV}^2 &
\end{eqnarray} 
for $P_{(5b)}$ states. Together with these above criteria, another criterion for these auxiliary parameters is the stability of the results with the variation of these parameters in their working intervals. Therefore, after fixing the intervals of these auxiliary parameters, we check the dependence of our results on them and depict their behaviors with the figures~\ref{gr:1}-\ref{gr:4} for both the full-charm and full-bottom pentaquark states. As is seen from these figures, these expectations are satisfied enough at the chosen working intervals.  
Having fixed the applicable intervals of the auxiliary parameters and by using all the inputs, the mass values and current coupling constants for these states  can be reported. The results are presented in Table~\ref{Table:II}. In the results, the errors arising from the errors of input parameters and uncertainties in the determination of the working intervals of the auxiliary parameters are also presented. 
\begin{table}[h!]
\begin{tabular}{|c|c|c|c|}
\hline\hline State &  $m~(\mathrm{MeV})$  & $\lambda~(\mathrm{GeV})^6$      \\ \hline\hline
 $P_{(5c)}$             &  $7628 \pm 112$  & $(9.81 \pm 2.97)\times10^{-2}$   \\
 \hline
 $P_{(5b)}$             &  $21982 \pm 144$ & $1.51 \pm 0.60$                 \\
\hline\hline
\end{tabular}%
\caption{: Masses and current coupling constants for $P_{(5c)}$ and $P_{(5b)}$ states.} \label{Table:II}
\end{table}

\section{Summary and conclusion}\label{IV} 

The number of members of exotic states is increasing day after day as a result of improvements in experimental techniques and analyses. Each new member added to this family increases the expectations of future observations of  such states and collects attention over these possible new states' investigations. With the motivations brought by the observed states, the ones with quark content different from the observed ones have been studied using various theoretical models to provide inputs or comparison grounds for future experimental findings. With the same motivation and the motivation brought by new observations of conventional or exotic states containing more numbers of heavy quarks, in the present work, we searched for possible pentaquark states composed of full charm or full bottom quarks and carrying spin-parity quantum numbers $J^P=\frac{3}{2}^-$. We predicted the masses for these states as $m = 7628 \pm 112$~MeV for full-charm pentaquark and $m = 21982 \pm 144$~MeV for full-bottom pentaquark.  In the literature, there are a few works investigating the full-heavy pentaquark states. The similar mass predictions in Ref.~\cite{Zhang:2020vpz} are $m=7.41^{+0.27}_{-0.31}$~GeV and $m=21.60^{+0.73}_{-0.22}$~GeV for full-charm and full-bottom states, respectively. In Ref.~\citep{Yan:2021glh}, the prediction  for the mass of the full-bottom pentaquark state with spin-parity $\frac{3}{2}^-$  is  $m=23748.2 \sim 23752.3$~MeV, on the other hand, they obtained no bound state for the charm counterpart of this state. The predictions in Ref.~\citep{An:2020jix} are $m=7863.6$~MeV and $m=23774.8$~MeV for full-charm and full-bottom pentaquarks with spin-parity $J^P=\frac{3}{2}^-$, respectively. Beside these, in Ref~\cite{An:2022fvs}, the analysis gave no stable full-heavy pentaquark system. The results of the present work are consistent with those of Ref.~\cite{Zhang:2020vpz} within the errors for both types of pentaquarks, on the other hand, they are smaller than the predictions presented in Refs.~\cite{Yan:2021glh,An:2020jix}. As is seen, there is a need to conduct more investigations about these states to gain a deeper understanding of them.     

Besides, since they are necessary for the calculation of the form factors that are used in the investigations of possible decay mechanisms of these states, we calculated the corresponding current coupling constants for these states.  

It is evident that further analyses of these states are necessary to gain a better understanding of their structure and properties. These results may be supported by other works through comparisons of their results or used in further investigations delving into the interaction mechanisms of these states.

\section*{ACKNOWLEDGEMENTS}
K. Azizi is thankful to Iran Science Elites Federation (Saramadan)
for the partial  financial support provided under the grant number ISEF/M/99171.



\end{document}